\documentclass[twocolumn,superscriptaddress,amsmath,amssymb]{revtex4}

\usepackage{graphicx}
\usepackage{dcolumn}
\usepackage{bm}

\def\beq{\begin{equation}}
\def\eeq#1{\label{#1}\end{equation}}
\def\beqa{\begin{eqnarray}}
\def\eeqa#1{\label{#1}\end{eqnarray}}

\def\sign{{\rm sign}}

\begin{document}


\title{A Multivariate Moran Process with Lotka-Volterra Phenomenology}
\author{Andrew E. Noble}
\affiliation{Department of Environmental Science and Policy, University of California, Davis, CA 95616, USA}%
\author{Alan Hastings}
\affiliation{Department of Environmental Science and Policy, University of California, Davis, CA 95616, USA}%
\author{William F. Fagan}
\affiliation{Department of Biology, University of Maryland, College Park, MD 20742, USA}

\date{\today}

\begin{abstract}

For a population with any given number of types, we construct a new multivariate Moran process with frequency-dependent selection and establish, analytically, a correspondence to equilibrium Lotka-Volterra phenomenology.  This correspondence, on the one hand, allows us to infer the phenomenology of our Moran process based on much simpler Lokta-Volterra phenomenology, and on the other, allows us to study Lotka-Volterra dynamics within the finite populations of a Moran process.  Applications to community ecology, population genetics, and evolutionary game theory are discussed.

\end{abstract}

\maketitle

The Moran process, originally formulated in the context of population genetics~\cite{Moran:1962}, has been applied to a wide range of systems in the biological sciences and statistical physics~\cite{Hofbauer:1998,Roca:2006p12387,Blythe:2007p12411,Claussen:2008p12391,Antal:2009}.  Multivariate Moran models prescribe mean-field stochastic dynamics for birth and death in a population with a fixed number of individuals and any given number of types~\cite{Claussen:2008p12391,Antal:2009}.  The deterministic limit yields replicator equations that model the evolution of type frequencies in large populations~\cite{Taylor:1978p12404,Schreiber:2001p12500,Traulsen:2005p12394}.  For either the multivariate Moran process or the corresponding replicator equations, types might be alleles, as in the population genetics of a single locus~\cite{Moran:1962,Gillespie:1994,Ewens:2004}; species, as in the neutral theory of community ecology~\cite{Caswell:1976p6359,PHubbell:2001p4284}; or strategies, as in evolutionary game theory~\cite{MaynardSmith:1982,Nowak:2004p12405,Nowak:2006,Nowak:2006p12410}.  Here, we build on a previous study of frequency-{\it independent} selection~\cite{Noble:2011p10042} to construct, for any given number of types, a new multivariate Moran model with frequency-{\it dependent} selection and demonstrate, analytically, that the corresponding replicator equations exhibit equilibrium Lotka-Volterra phenomenology.  The results presented here are distinct from the well-known equivalence of $S$ replicator equations and $S-1$ Lotka-Volterra equations~\cite{Hofbauer:1998,Tokita:2004p12413}.  Furthermore, our choice of frequency-dependent selection generates $S$ replicator equations with the equilibrium phenomenology of $S$ Lotka-Volterra equations.  ``Lotka-Volterra equations" in this context does not refer to the special case of neutrally stable, two-species, predator-prey dynamics, but rather to the general case of mean-field deterministic dynamics for a population with any number of well-mixed types and linearized density-dependent growth rates~\cite{Volterra:1931}.  These generalized Lotka-Volterra equations are frequently used to model competitive dynamics in ecological communities~\cite{MMay:1974p9022}.  Indeed, the two-species case may be the simplest dynamical system to exhibit the niche mechanism for stabilizing coexistence, in which (i) demographic rates must vary among species, (ii) abundances for each species must increase when rare, and (iii) {\it intra}specific competition must exceed {\it inter}specific competition~\cite{Chesson:2000p5531}.  By embedding Lotka-Volterra dynamics within a multivariate Moran process, we obtain a theoretical framework in which equilibrium Moran phenomenology may be inferred from much simpler Lotka-Volterra phenomenology, and Lotka-Volterra dynamics may be studied within the finite populations of a Moran process.  This framework offers new insights on community ecology, population genetics, and evolutionary game theory.

In a multivariate Moran process for $S$ types and $N$ individuals, the allowed states are vectors of nonnegative integers, $\vec{n}=(n_1,\dots,n_S)$, such that $0 \le n_i \le N$ for each $i$ and $\sum_{i=1}^Sn_i=N$.  The allowed transitions are a single death event immediately followed by a single birth event in order to maintain a total of $N$ individuals.  In the absence of selection, i.e.~the neutral limit, per capita rates of birth and death are equivalent.  If we ignore mutation and migration, the neutral transition rate for a death event in Type $i$ immediately followed by a birth event in Type $j$ is simply
\beq
T_{ij\vec{n}} \,=\,\frac{n_i}{N}\left(\frac{n_j}{N-1}\right),
\eeq{snmt}
where only $N-1$ individuals are present after the death of $i$ and prior to the birth of $j$.  Dynamics are governed by a multivariate master equation that we can write as~\cite{Noble:2011p10042}
\beq
\frac{dP_{\vec{n}}}{d\tau}\,=\,\sum_{i=1}^S\sum_{j=1}^S\left(T_{ij\vec{n}+\vec{e}_i-\vec{e}_j}P_{\vec{n}+\vec{e}_i-\vec{e}_j}-T_{ji\vec{n}}P_{\vec{n}} \right) \Theta_{ij\vec{n}},
\eeq{master}
where $P_{\vec{n}}$ is the probability of state $\vec{n}$, $\vec{e}_i$ is a unit vector, and $\tau$ is a dimensionless measure of time.  The $\Theta_{ij\vec{n}}\equiv \Theta(N-(n_i+1))\Theta(n_j-1)$, where $\Theta(x)=0$, for $x<0$, and $1$ otherwise, eliminate transitions to non-allowed states.  The stochastic process contains $S$ absorbing states where a single species dominates such that $n_i=N$ for some $i$.  The more familiar univariate Moran process is obtained from the marginal dynamics of Eq.~\ref{master}~\cite{Noble:2011p10042}.

For $S$ types, the Lotka-Volterra equations can be written as
\beq
\frac{dx_i}{d\tau}\,=\,x_if_i(\vec{x}),
\eeq{lv}
where $x_i=x_i(\tau)$ is the density of Type $i$ and
\beq
f_i(\vec{x})\,=\,r_i-\sum_{j=1}^S a_{ij}x_j,
\eeq{lvf}
is a density-dependent growth rate parameterized by $r_i$, the intrinsic growth rate of Type $i$ in the absence of all others, and $a_{ij}$, the additive per capita impact of Type $j$ on the growth rate of Type $i$.  We focus for now on {\it competition} among types such that $r_i>0$ and $a_{ij}>0$ for all $i$ and $j$.  In community ecology, the $a_{ii}$ are referred to as ``intraspecific competition strengths" and the $a_{ij}$, with $i\ne j$, as ``interspecific competition strengths".   Without loss of generality, we can re-scale the Lotka-Volterra equations such that 
\beq
r^\prime_i\,\equiv\,\frac{r_i}{\vert\sum_{k=1}^Sr_k\vert}, \quad a^\prime_{ij}\,\equiv\,\sign\left(\sum_{k=1}^Sr_k\right) \frac{a_{ij}}{\sum_{k=1}^Sa_{kj}}, \nonumber
\eeq{}
\vskip-0.45cm
\beq
x^\prime_i\,\equiv\,\frac{\sum_{k=1}^Sa_{ki}}{\sum_{k=1}^Sr_k}x_i, \quad \tau^\prime\,\equiv\,\tau\left\vert\sum_{k=1}^Sr_k\right\vert, 
\eeq{rescale}
and if a coexisting fixed point, $\vec{x}^{\prime*}$, exists, with $x_i^{\prime*}>0$ for all $i$, then 
\beqa
\sum_{i=1}^Sx^{\prime*}_i\,=\,1.
\eeqa{} 

With these preliminaries, we first construct a new multivariate Moran process and then establish a correspondence to equilibrium Lotka-Volterra phenomenology.  For the construction phase, we seek frequency-dependent selection coefficients similar in form to the density-dependent growth rates of the Lotka-Volterra equations but remaining nonnegative for all states and parameter values.  Towards this end, the growth rates of a Ricker model~\cite{Ricker:1954} inspire our choice of frequency-dependent selection coefficients
\beq
w_{i\vec{n}}\equiv \exp\left(r^\prime_i-\sum_{k=1}^S a^\prime_{ik}\frac{n_k}{N}\right),
\eeq{introwi}
and we expand on Eq.~\ref{snmt} to impose the transition rates
\beqa
T_{ij\vec{n}} \,=\,\frac{n_i}{N}\left(\frac{w_{j\vec{n}-\vec{e}_i}n_j}{\sum_{k=1}^Sw_{k\vec{n}-\vec{e}_i}n_k-w_{i\vec{n}-\vec{e}_i}}\right),
\eeqa{snnmt}
where various subtractions account for the death of Type $i$ prior to the birth of Type $j$.  In the context of population genetics, the $w_{i\vec{n}}$ are reproductive fitnesses, and the multivariate Moran process of Eq.~\ref{master} with the transition rates of Eq.~\ref{snnmt} provides a framework for multiallelic, frequency-dependent selection that is not limited by the usual assumptions of weak selection or symmetry under the exchange of alleles~\cite{Muirhead:2009p12401}.  Symmetry under exchangeability only obtains in the neutral limit where Eq.~\ref{snnmt} reduces to Eq.~\ref{snmt} for $r^\prime_i=r^\prime_j$ and $a^\prime_{ij}=1/S$ for all $i$ and $j$.   

\begin{figure*}[t!]
\centerline{
\includegraphics[width=5.75cm]{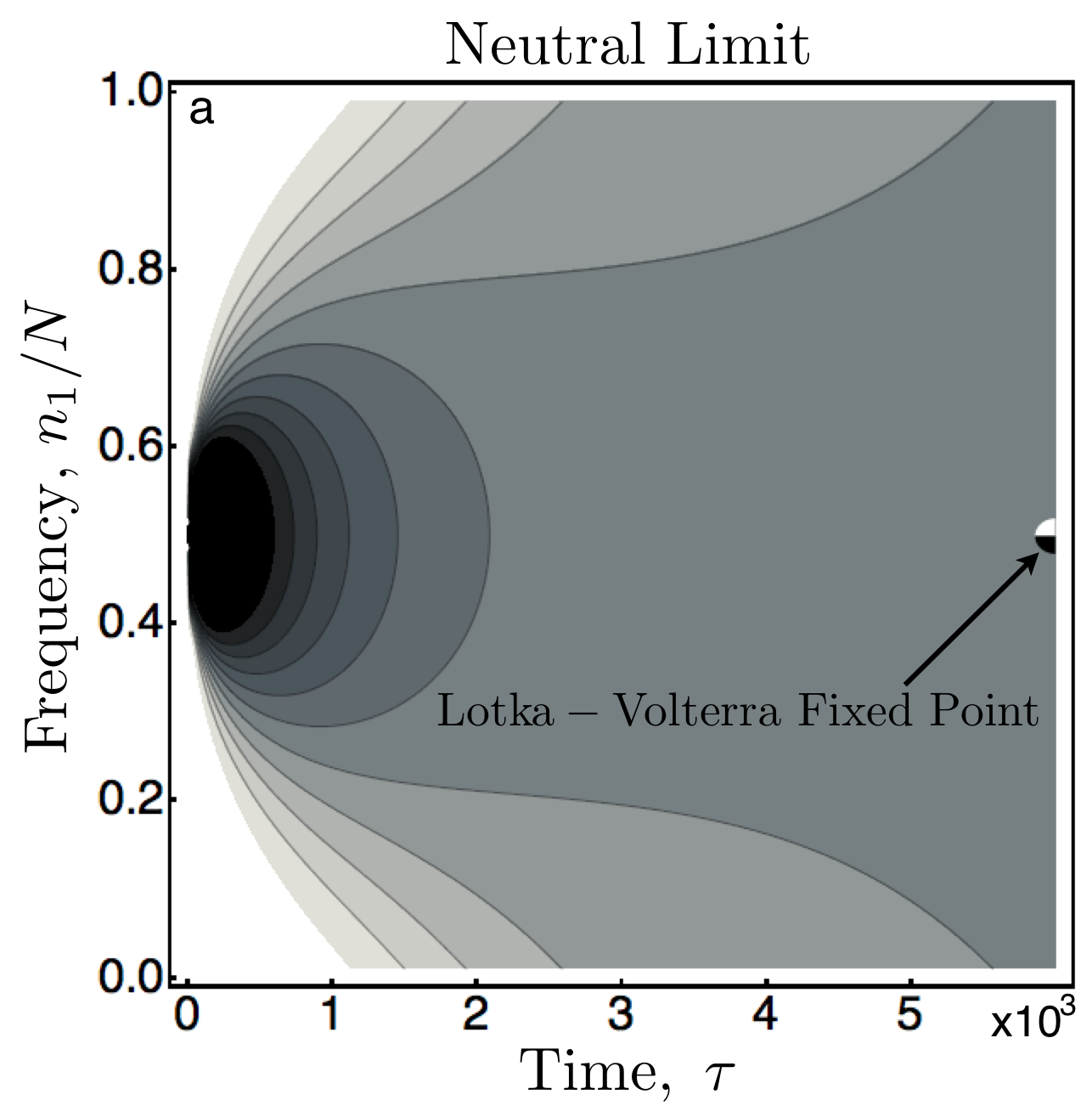}
\hskip0.00cm
\includegraphics[width=4.95cm]{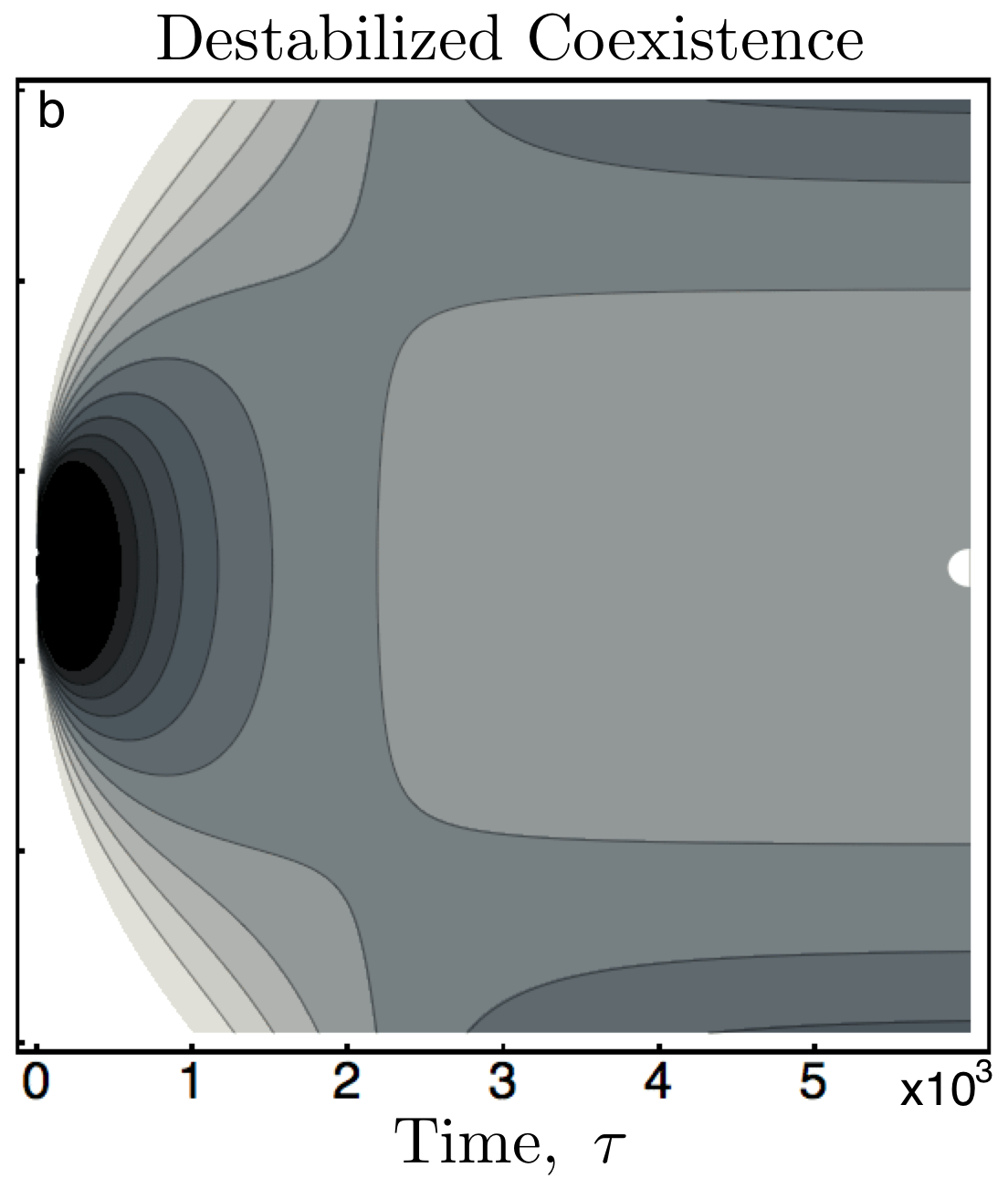}
\hskip0.00cm
\includegraphics[width=6.27cm]{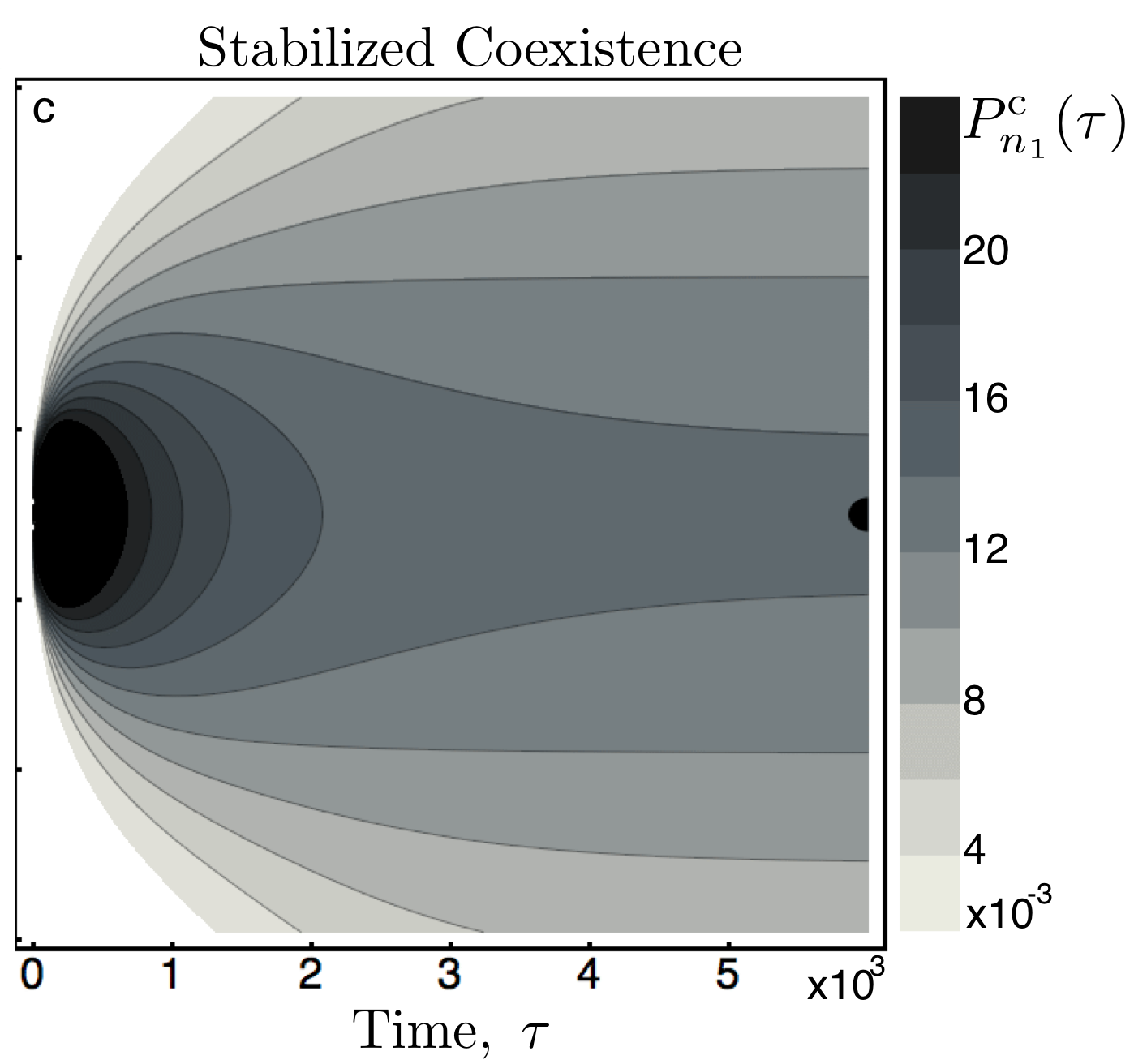}
}
\vskip-0.425cm
\caption{Plots of the integrated marginal distribution for Type 1 in a two-type population after conditioning against extinction and dominance, $P^{\rm c}_{n_1}(\tau) = P_{n_1}(\tau)/(1-P_0(\tau)-P_N(\tau))$.  For all plots, $N=100$, $r^\prime_1=r^\prime_2=0.5$, and the initial condition is $P^{\rm c}_{N/2}(\tau=0)=1$.  The panels correspond to scenarios of (a) neutrality ($a^\prime_{11}=a^\prime_{21}=a^\prime_{22}=a^\prime_{12}=0.5$), (b) interspecific exceeding intraspecific competition ($a^\prime_{21},a^\prime_{12}=0.52>a^\prime_{11},a^\prime_{22}=0.48$), and (c) intraspecific exceeding interspecific competition ($a^\prime_{21},a^\prime_{12}=0.48<a^\prime_{11},a^\prime_{22}=0.52$).  At long times, the conditional probabilities approach quasi-stationary distributions with shapes anticipated by the stability of the coexisting Lotka-Volterra fixed point which is (a) neutrally stable (half-filled semicircle), (b) unstable (empty semicircle), and (c) stable (filled semicircle).}
\label{msi}
\vskip-0.5cm
\end{figure*}

\begin{figure}[]
\centerline{
\includegraphics[width=6.cm]{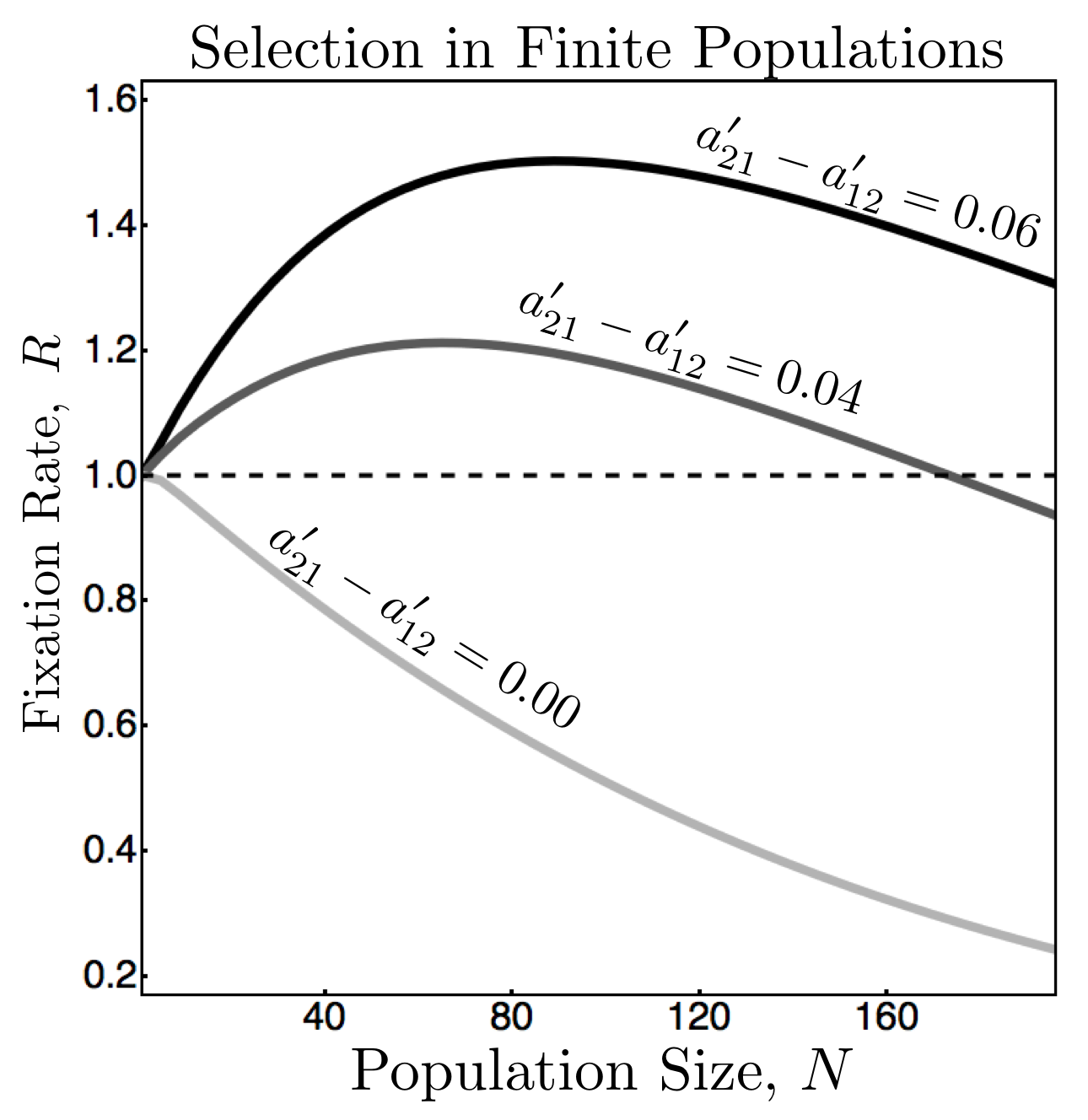}
}
\vskip-0.5cm
\caption{A plot of fixation rate, $R$, as a function of the number of individuals, $N$, in a two-type population where $r^\prime_1=r^\prime_2=0.5$,  $a^\prime_{12}=0.52$, and $a^\prime_{21}=0.52$ (light grey), $a^\prime_{21}=0.56$ (dark grey), or $a^\prime_{21}=0.58$ (black), with $a^\prime_{11}=1-a^\prime_{21}$ and $a^\prime_{22}=1-a^\prime_{12}$.  Type 1 is always excluded as $N\rightarrow\infty$, but for sufficiently small $N$, the $a^\prime_{21}>a^\prime_{12}$ asymmetry is sufficiently strong that frequency-dependent selection yields fixation rates above the neutral limit (dashed line).}
\label{rvsj}
\vskip-0.5cm
\end{figure}

The deterministic limit of Eq.~\ref{master}, with the transition rates of Eq.~\ref{snnmt}, is given by
\beq
\frac{dn_i}{d\tau^{\prime\prime}}\,=\,N\sum_{j=1}\left(T_{ji\vec{n}}-T_{ij\vec{n}}\right) 
\eeq{deterlimitv0}
where $\tau^{\prime\prime}\equiv\tau/N$.  A simplification is obtained in the case of weak asymmetries where
\beq
a^\prime_{ij}<<\sum_{k=1}^S a^\prime_{ik}n_k, \quad w_{i\vec{n}}<<\sum_{k=1}^S w_{k\vec{n}}n_k,
\eeq{asymmreq}
for every $i$, $j$, and $\vec{n}$.   ``Weak asymmetries" in our multivariate Moran process are similar to ``weak selection" in a univariate Moran process~\cite{Hofbauer:1998,Ewens:2004}, but due to the weighted sums over $n_k$ that appear in Eq.~\ref{asymmreq}, our asymmetry requirements depend on population size, $N$, as highlighted below in the discussion of Fig.~\ref{rvsj}.  If we assume sufficiently weak asymmetries, and if we transform variables from densities, $n_i$, to frequencies, $p_i\equiv n_i/N$, Eq.~\ref{deterlimitv0} can be approximated by the replicator equation
\beq
\frac{dp_i}{d\tau^{\prime\prime}}\,\sim\,p_i \left(c_i(\vec{p}) - 1\right),
\eeq{deterlimit}
where the replicator fitness is 
\beq
c_{i}(\vec{p})\,\equiv\,\frac{w_{i}(\vec{p})}{\sum_{k=1}^S p_k w_{k}(\vec{p})},
\eeq{cdef}
and $w_{i}(\vec{p})$ is defined by Eq.~\ref{introwi} upon substituting $p_k N$ for $n_k$.  The mean replicator fitness is unity at all times, and therefore, never decreases - a standard requirement of replicator dynamics~\cite{Hofbauer:1998}.

To establish a correspondence to equilibrium Lotka-Volterra phenomenology, we note that a coexisting fixed point for the Lotka-Volterra system of Eq.~\ref{lv} is also a coexisting fixed point for the replicator system of Eq.~\ref{deterlimit} 
\beq
\vec{p}^*\,=\,\vec{x}^{\prime*}.
\eeq{}
But in addition to this identity, we find a correspondence in stability:  if the matrix with elements given by $a^\prime_{ij}+a^\prime_{ji}$ is positive definite, then the coexisting fixed point is globally stable in the Lotka-Volterra system and at least locally stable in the replicator system.  The proof for any given number of types employs the well-known Lyapunov function for Lotka-Volterra dynamics~\cite{Getz:1975,Goh:1977p12466} (see Supplemental Material).  In the special case of $S=2$ competition, if $r^\prime_1a^\prime_{21}<r^\prime_2a^\prime_{11}$ and $r^\prime_2a^\prime_{12}<r^\prime_1a^\prime_{22}$, then the replicator and Lotka-Volterra equations share a stable coexisting fixed point and each type can invade when rare.  The latter inequalities, which imply that $a^\prime_{12}a^\prime_{21}<a^\prime_{11}a^\prime_{22}$, quantify the three requirements of niche theory that were highlighted in the introduction.  

To illustrate how Lotka-Volterra phenomenology can facilitate inference on our multivariate Moran process, we consider a two-type population where Eq.~\ref{master} can be re-written as a univariate birth-death process~\cite{Noble:2011p10042,VanKampen:2001} with rates of gain and loss given by
\beqa
b_{i\vec{n}} &=& \frac{N-n_1}{N}\left(\frac{w_{1\vec{n}-\vec{e}_2}n_1}{w_{1\vec{n}-\vec{e}_2}n_1+w_{2\vec{n}-\vec{e}_2}(N-n_1-1)}\right), \nonumber \\
d_{i\vec{n}} &=& \frac{n_1}{N}\left(\frac{w_{2\vec{n}-\vec{e}_1}(N-n_1)}{w_{1\vec{n}-\vec{e}_1}(n_1-1)+w_{2\vec{n}-\vec{e}_1}(N-n_1)}\right), \nonumber \\
\eeqa{}
and $\vec{n}=(n_1,N-n_1)$.  Starting from a known initial abundance, Fig.~\ref{msi} integrates the marginal distribution of Type 1 conditioned against extinction and dominance.  We consider three cases:  a) the neutral limit, b) destabilized coexistence due to interspecific exceeding intraspecific competition, and c) stabilized coexistence due to intraspecific exceeding interspecific competition.  The quasi-stationary distribution emerging at long times is (a) flat, (b) peaked at extinction and dominance, or (c) peaked at coexistence.  In the corresponding Lotka-Volterra equations, the stability of the coexisting fixed point at $n_1/N=0.5$ allows us to infer the shape of the quasi-stationary distribution with (a) neutral stability linked to a flat distribution, (b) instability linked to a local minimum, and (c) stability linked to a local maximum~\cite{VanKampen:2001}.  As $N$ becomes large, fluctuations become rare and the peaks in Figs.~\ref{msi}b and \ref{msi}c approach delta functions.  If we interpret Fig.~\ref{msi} in the context of diploid population genetics, we find that interspecific exceeding intraspecific competition yields a scenario of ``underdominance" (Fig.~\ref{msi}b), in which selection favors homozygous over heterozygous populations, and intraspecific exceeding interspecific competition yields a scenario of ``overdominance" (Fig.~\ref{msi}c), in which selection favors heterozygosity~\cite{Gillespie:1994}.  The Supplemental Material provides predator-prey and multi-type examples to further illustrate the use of Lotka-Volterra dynamics to anticipate the phenomenology of our multivariate Moran process.  For the multi-type example, Moran dynamics are simulated with the Gillespie algorithm~\cite{Gillespie:1977p12590} as implemented in the Python package StoMPy~\cite{Maarleveld:2010}.

But we also need to consider cases where the assumptions of Eq.~\ref{asymmreq} break down, and the correspondence between Moran process and Lotka-Volterra system no longer holds.  Fortunately, in this regime, the Moran process remains valid for strong asymmetries and allows us to study Lotka-Volterra dynamics within finite populations.  As an example, consider fixation rates, or the expected rate at which individuals of a given type grow to dominate a population given a fixed rate of repeated introduction.  In a two-type population, the fixation rate of Type 1 is given by~\cite{Nowak:2006}
\beq
R\,=\,N\left(1+\sum_{k=1}^{N-1}\prod_{i=1}^k\frac{d_{i\vec{n}}}{b_{i\vec{n}}}\right)^{-1},
\eeq{}
in our nondimensionalized time units.  For large populations that satisfy the weak asymmetry requirements of Eq.~\ref{asymmreq}, fixation rates are vanishingly small when $r^\prime_2a^\prime_{12}>r^\prime_1a^\prime_{22}$, and as expected from Lotka-Volterra dynamics, Type 1 cannot invade Type 2.  However, in the stochastic dynamics of small populations with strong asymmetries, the introduction of a single individual of Type 1 can disproportionately impact the fitness of Type 2 if $a^\prime_{21}>a^\prime_{12}$.  For three different values of $a^\prime_{21}-a^\prime_{12}$, Fig.~\ref{rvsj} plots fixation rates against population sizes in a scenario where $r^\prime_2a^\prime_{12}>r^\prime_1a^\prime_{22}$.  When $a^\prime_{21}-a^\prime_{12}>0$, fixation rates lie above the neutral limit of $R=1$ for sufficiently small populations.  Therefore, in our multivariate Moran process, strong asymmetries in finite populations can favor invasion even when the corresponding Lokta-Volterra system for an infinite population predicts exclusion.   

We conclude by discussing two applications of our framework for merging Moran and Lotka-Volterra dynamics.  In community ecology, the past decade has witnessed an enormous interest in predicting large-scale abundance distributions based on neutral models for individual interactions.  The foundational work of Caswell~\cite{Caswell:1976p6359} and Hubbell~\cite{PHubbell:2001p4284} applied the neutral Moran models of population genetics to competitive ecological communities, such as the canopy trees in a tropical forest where light-gaps are quickly filled and the total number of individuals may be approximated as a constant.  For these models, types correspond to species and genetic variation within each species is ignored.  Efforts to determine the relative importance of deterministic niche dynamics and stochastic neutral dynamics in structuring ecological communities have been limited by the lack of a common analytical framework.  Fortunately, by adding mutation or migration to the transition rates of our multivariate Moran process, where selection occurs on the level of species rather than alleles in this context, we can unify Hubbell's formulation of neutral theory~\cite{PHubbell:2001p4284} with niche theory~\cite{Chesson:2000p5531}.  In particular, with the addition of a small mutation rate, the deterministic limit of our multivariate Moran process yields a replicator-mutator equation~\cite{Nowak:2006,Nowak:2006p12410} exhibiting equilibrium Lotka-Volterra phenomenology up to small corrections.  This non-neutral framework, in which niche stabilization can delay extinction, offers a simple solution to Hubbell's problem of short species lifetimes~\cite{PHubbell:2001p4284}.  

Merging Moran and Lotka-Volterra dynamics is also relevant to evolutionary game theory.  Indeed, Fig~\ref{rvsj} is reminiscent of results in Nowak {\it et al.}~\cite{Nowak:2004p12397} for a univariate Moran process with frequency-dependent selection that models evolutionary games in finite populations.  Their definition of reproductive fitness can be mapped onto a first-order expansion of Eq.~\ref{introwi} if we now assume that $r_i<0$ and $a_{ij}<0$ for all $i$ and $j$.  In this regime, the equilibrium phenomenology of our re-scaled Lotka-Volterra and replicator equations matches the usual expectations for two-strategy games, such as the Prisoner's Dilemma~\cite{Axelrod:1981p12407} and Repeated Prisoner's Dilemma~\cite{Nowak:2004p12397,Taylor:2004p12398}, given $r_1=r_2$ and payoff matrix 
\beq
\bordermatrix{
                                         &{\rm Strategy~1} & {\rm Strategy~2}\cr
               {\rm Strategy~1}& -a_{11} & -a_{12}\cr
               {\rm Strategy~2}& -a_{21} & -a_{22}
               }.
\eeq{}  
\\
Our multivariate Moran process with frequency-dependent selection provides a general framework for modeling the stochastic dynamics of evolutionary games with any given number of strategies in a finite population. 

We are grateful for productive conversations with Bill Creskoe, Jessica Green, Matt Holland, Marcel Holyoak, Tim Keitt, Sivan Leviyang, Timo Maarleveld, Patrick Phillips, Annette Ostling, Bruce Rannala, Sahotra Sarkar, Sebastian Schreiber, and an annonymous reviewer.  Our work is partially supported by the NSF through their Emerging Frontiers grant (0827460) to A.~H.~and by the James S. McDonnell Foundation through their Studying Complex Systems grant (220020138) to W.~F.~F.

\clearpage

\section*{Supplemental Material}

\subsubsection*{A correspondence in stability between the Lotka-Volterra and replicator systems}

For a system of $S$ Lotka-Volterra equations, as given by Eq.~3 of the main text, the well-known Lyapunov function can be written as
\beq
V(\vec{x}^\prime)\,=\,\sum_{i=1}^S\left(x^\prime_i-x^{\prime*}_i-x^{\prime*}_i\log x^\prime_i/x^{\prime*}_i \right),
\tag{SM.1}
\eeq{lyap}
where $x^\prime_i>0$ for all $i$, $\vec{x}^{\prime*}$ is a unique global minimum, and the time evolution is~\cite{Getz:1975:cite2,Goh:1977p12466:cite2}
\beq
\frac{dV(\vec{x}^\prime)}{d\tau}\,=\,-\frac{1}{2}\sum_{i=1}^S\left(x^\prime_i-x^{\prime*}_i\right)\left(a^\prime_{ij}+a^\prime_{ji}\right)\left(x^\prime_j-x^{\prime*}_j\right).
\tag{SM.2}
\eeq{dvdt}    
Under replicator dynamics, the right-hand-side of Eq.~SM.2, with $x_i^\prime\rightarrow p_i$, is the leading term in an expansion of $dV(\vec{p})/d\tau^{\prime\prime}$ about $\vec{p}^*$.  Therefore, $\vec{x}^{\prime*}$ is globally stable, and $\vec{p}^*$ is at least locally stable, if $a^\prime_{ij}+a^\prime_{ji}$ is positive definite.  

\subsubsection*{A predator-prey example}

The main text restricts attention to Lotka-Volterra dynamics in which all the $r_i$ and $a_{ij}$ share the same sign, but the rescaling of Eq.~5 can be applied more generally.  Of course, the rescaling of Eq.~5 fails when parameters are tuned such that $\sum_{k=1}^Sr_k=0$ or $\sum_{k=1}^Sa_{ki}=0$ for some $i$, but this problem occurs with vanishing probability when $r_i$ and $a_{ij}$ are empirically estimated.  Fig.~\ref{pp} illustrates our ability, in a predator-prey system, to infer the phenomenology of our multivariate Moran process from the corresponding Lotka-Volterra equations.  A stable fixed point of the Lotka-Volterra system at $n_1/N=0.5$ anticipates the mean value of the prey's marginal stationary distribution.  

\setcounter{figure}{0}
\renewcommand{\thefigure}{SM.\arabic{figure}}

\begin{figure}[h!]
\centerline{
\hskip1.15cm \includegraphics[width=6.25cm]{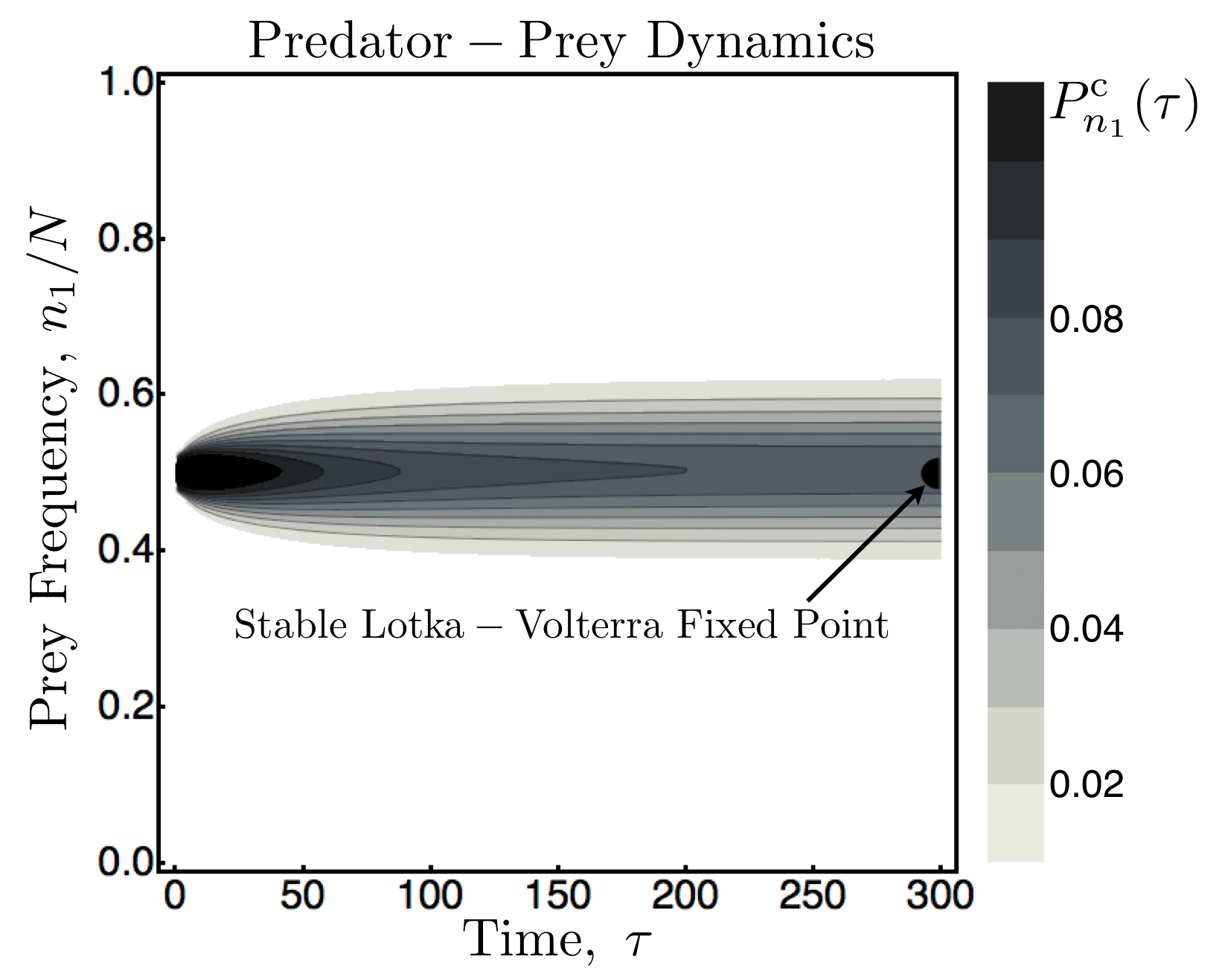}
}
\vskip-0.425cm
\caption{The integrated marginal distribution for the prey in a predator-prey population after conditioning against extinction and dominance, $P^{\rm c}_{n_1}(\tau) = P_{n_1}(\tau)/(1-P_0(\tau)-P_N(\tau))$, for system size $N=100$, initial condition $P^{\rm c}_{N/2}(\tau=0)=1$, intrinsic growth rates $r^\prime_1=1.25$, $r^\prime_2=-0.25$, and interaction strengths $a^\prime_{11}=2.00$, $a^\prime_{21}=-1.00$, $a^\prime_{22}=a^\prime_{12}=0.50$.  At long times, the conditional probability approaches a quasi-stationary distribution with a local maximum anticipated by the stable fixed point of the Lotka-Volterra dynamics.}
\label{pp}
\vskip-0.5cm
\end{figure}

\begin{figure*}[t!]
\centerline{
\includegraphics[width=5.8cm]{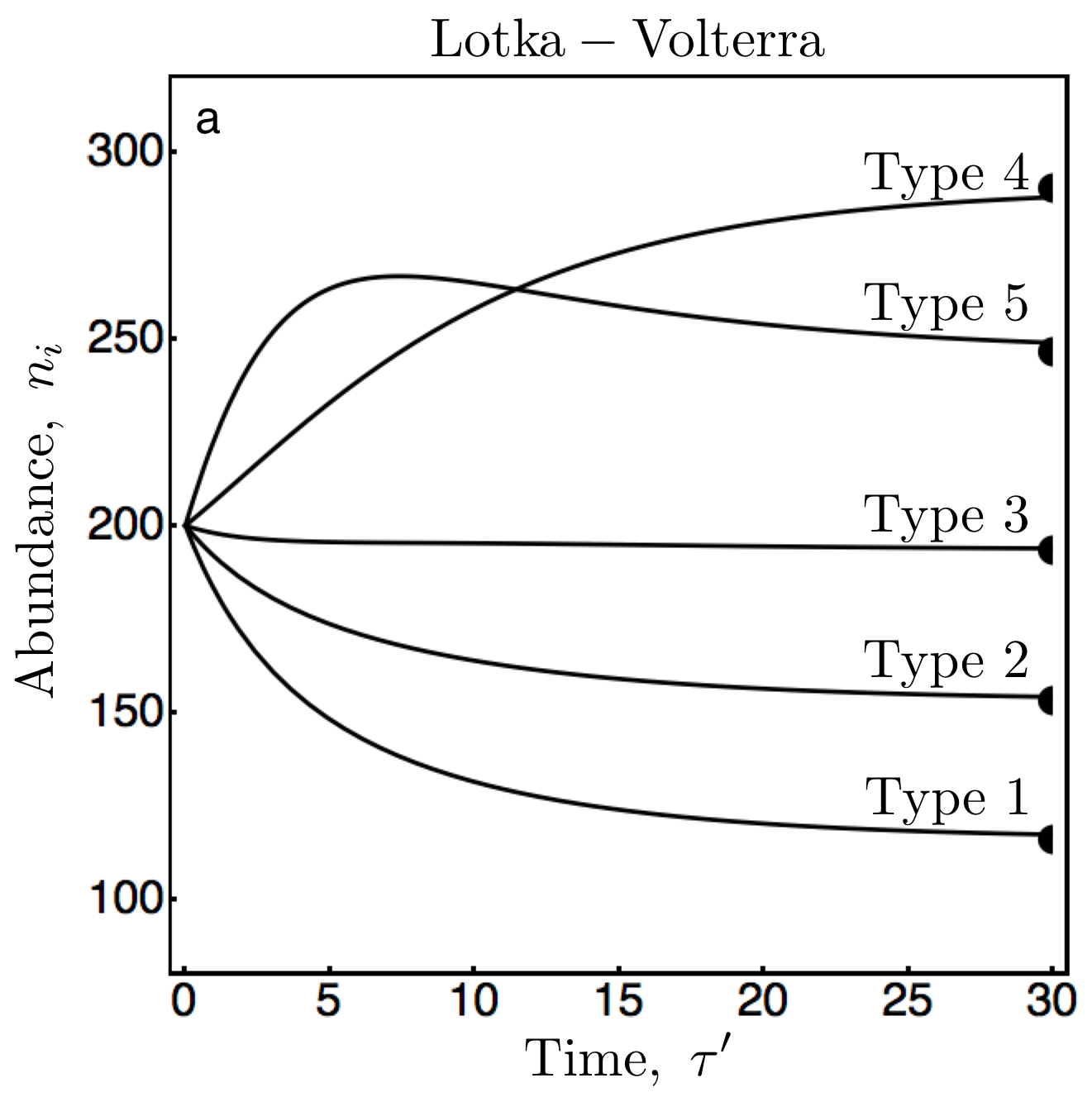}
\includegraphics[width=5.cm]{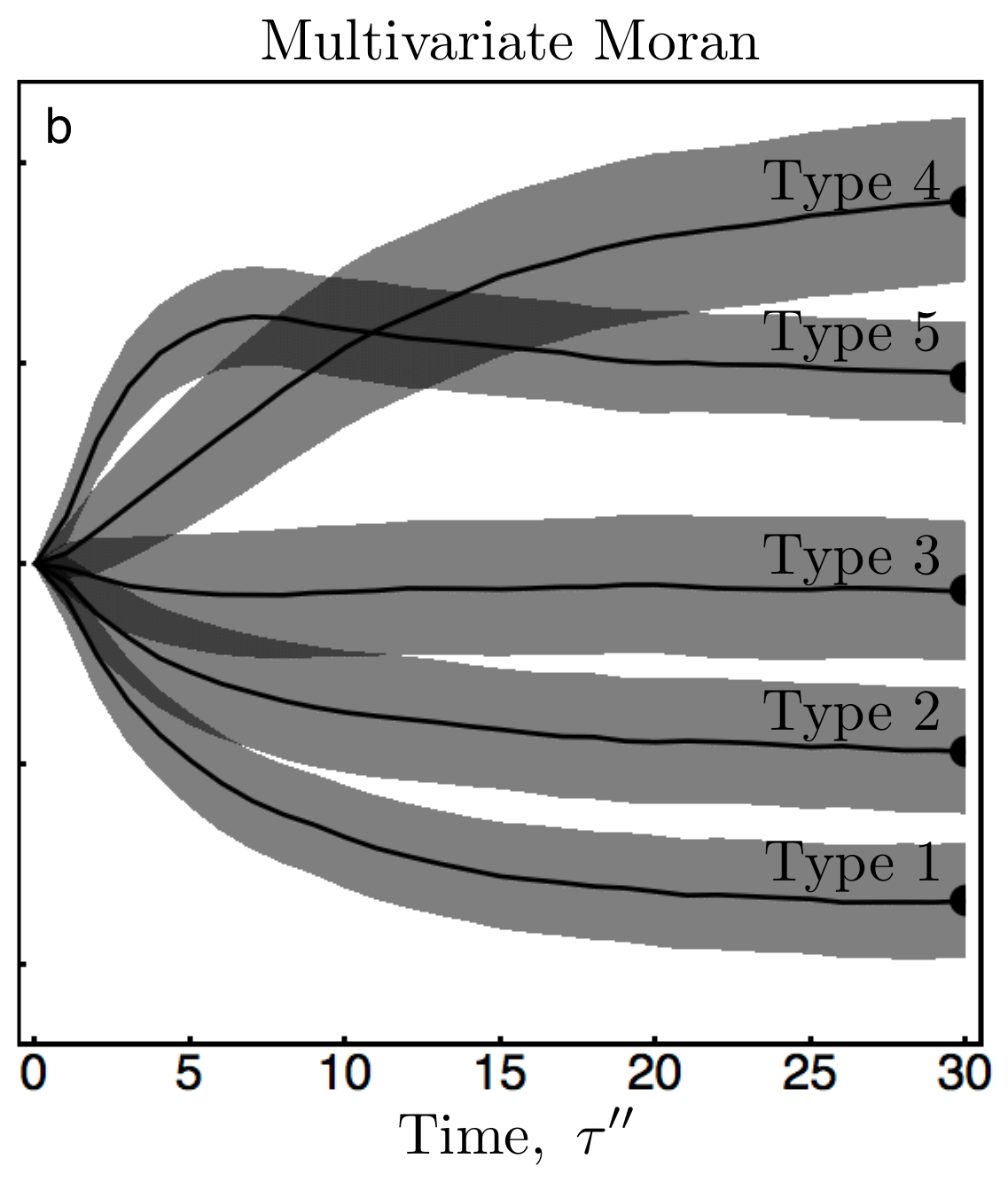}
}
\vskip-0.425cm
\caption{A comparison of Lotka-Volterra dynamics with the corresponding multivariate Moran process in our multi-type example for $S=5$ and $J=1000$.  All types have an initial abundance of 200 in both cases.  Panel (a) displays deterministic trajectories in the Lotka-Volterra system.  Panel (b) plots means (black lines) and standard deviations (light gray regions) for integrated marginal distributions in the multivariate Moran process after conditioning against extinction or dominance by any single type.  Dark gray regions indicate overlap in the standard deviations.  At long-times, mean abundances of marginal stationary distributions in the multivariate Moran process converge to stable equilibrium abundances of the corresponding Lotka-Volterra dynamics (filled semicircles).}
\label{mt}
\vskip-0.5cm
\end{figure*}

\subsubsection*{A multi-type example}

The main text establishes an analytical correspondence between our multivariate Moran process and equilibrium Lotka-Volterra phenomenology for any given number of types.  Here, we provide a multi-type example in which the equilibrium phenomenology of our Moran process is anticipated by the corresponding Lotka-Volterra dynamics.  We construct a consumer-resource system with $S=5$ types and $J=1000$ individuals where the fifth type (consumer) benefits from the first four types (resources).  Parameters are chosen such that, after the re-scalings of Eq.~5, the interaction matrix, with elements $a_{ij}^\prime$, is 
\beq
a^\prime\,=\,\left(
\begin{array}{ccccc}
1.333  & 0.211  & 0.222  & 0.222  & 0.022 \\
0.222  & 1.053  & 0.222  & 0.222  & 0.022 \\
0.222  & 0.211  & 0.889  & 0.222  & 0.022 \\
0.222  & 0.211  & 0.222  & 0.667  & 0.022 \\
-1.000  & -0.684  & -0.556  & -0.333  & 0.914   
\end{array} \nonumber
\right),
\eeq{}
the vector of intrinsic growth rates, with elements $r_i^\prime$, is 
\beq
r^\prime\,=\,(0.300,~0.300,~0.300,~0.300,~-0.200), \nonumber
\eeq{}
and all numerical values have been rounded to the nearest one-thousandth.  Trajectories of the re-scaled Lotka-Volterra system are plotted in Fig.~\ref{mt}a with filled semicircles indicating equilibrium abundances that are globally stable due to $a^\prime+a^{\prime T}$ being positive definite.  The dynamics of our Moran process were simulated with the Gillespie algorithm~\cite{Gillespie:1977p12590:cite2} as implemented in the Python package StoMPy~\cite{Maarleveld:2010:cite2}.  We ran 2000 stochastic trajectories and calculated statistics over time bins of width $\tau^{\prime\prime}$=1.  Fig.~\ref{mt}b plots means and standard deviations for integrated marginal distributions conditioned against extinction or dominance by any single type.  Clearly, at long-times, mean abundances of marginal stationary distributions in the multivariate Moran process converge to stable equilibrium abundances of the corresponding Lotka-Volterra dynamics.  Even over short and intermediate time scales, the dynamics are remarkably similar.  While simulations of our multivariate Moran process become computationally intensive as the number of types becomes large, the Lotka-Volterra system remains relatively easy to integrate.

\end{document}